\documentclass[preprint]{imsart}

\RequirePackage[OT1]{fontenc}
\usepackage{amsthm,amsmath,natbib}

\startlocaldefs

\numberwithin{equation}{section}
\theoremstyle{plain}

\usepackage{amssymb}
\usepackage{graphicx}
\usepackage{url}
\usepackage{xcolor}
\usepackage{verbatim}
\usepackage{multirow}
\usepackage{tikz}
\usepackage{bbold}
\usepackage{mathtools}
\usepackage{float}

\usetikzlibrary{shapes,decorations,arrows,calc,arrows.meta,fit,positioning}

\newcommand\independent{\protect\mathpalette{\protect\independenT}{\perp}}
\def\independenT#1#2{\mathrel{\rlap{$#1#2$}\mkern2mu{#1#2}}}

\usepackage[normalem]{ulem}

\endlocaldefs

\begin{document}

\begin{frontmatter}

\title{The Role of Exchangeability in Causal Inference}
\runtitle{Exchangeability in Causal Inference}


\begin{aug}
\author[1]{\fnms{Olli} \snm{Saarela}\corref{}\ead[label=e1]{olli.saarela@utoronto.ca}},
\author[2]{\fnms{David A.} \snm{Stephens}\ead[label=e2]{d.stephens@math.mcgill.ca}} \and
\author[3]{\fnms{Erica E. M.} \snm{Moodie}\ead[label=e3]{erica.moodie@mcgill.ca}}

\runauthor{O. Saarela et al.}

\address[1]{Dalla Lana School of Public Health, University of Toronto. \printead{e1}.}

\address[2]{Department of Mathematics and Statistics, McGill University}

\address[3]{Department of Epidemiology and Biostatistics, McGill University}
\end{aug}

\begin{abstract}
Though the notion of exchangeability has been discussed in the causal inference literature under various guises, it has rarely taken its original meaning as a symmetry property of probability distributions. As this property is a standard component of Bayesian inference, we argue that in Bayesian causal inference it is natural to link the causal model, including the notion of confounding and definition of causal contrasts of interest, to the concept of exchangeability. Here we propose a probabilistic between-group exchangeability property as an identifying condition for causal effects, relate it to alternative conditions for unconfounded inferences (commonly stated using potential outcomes) and define causal contrasts in the presence of exchangeability in terms of posterior predictive expectations for further exchangeable units. While our main focus is on a point treatment setting, we also investigate how this reasoning carries over to longitudinal settings.
\end{abstract}

\begin{keyword}
\kwd{Bayesian inference}
\kwd{Causal inference}
\kwd{Confounding}
\kwd{Exchangeability}
\kwd{Posterior predictive inference}
\end{keyword}

\end{frontmatter}

\section{Introduction}\label{section:introduction}

The concept of exchangeability has profound philosophical meaning in Bayesian statistics.  Recall that an infinite sequence of
observable random variables $(Y_i)_{i=1}^{\infty}$ is \textit{exchangeable} if, for all finite $n$,
\begin{align}\label{eq:ex}
\MoveEqLeft[1] \Pr(Y_1 = y_1,\ldots,Y_n=y_n) \\
&= \Pr(Y_1 = y_{\rho(1)},\ldots,Y_n=y_{\rho(n)}) \nonumber,
\end{align}
or $(Y_1,\ldots,Y_n) \stackrel{\textrm{d}}{=} (Y_{\rho(1)},\ldots,Y_{\rho(n)})$, for any permutation $\rho(.)$ of the indices. This simple probabilistic definition plays a central, even totemic, role in Bayesian inference; it leads to the definition of `parameters' as functions of infinite sequences of observable quantities through de Finetti's representation theorem (\citealp{de1929funzione}; a review of the original work is provided for example by \citealp{von1989finetti}). This further facilitates probability statements on future, unobserved quantities based on information contained in observed data, and justifies the use of the posterior distribution as the basis for statistical inference \citep[e.g.,][p. 173]{bernardo:1994}. In recent years, the term `exchangeability', or `conditional exchangeability', has been increasingly used in the field of causal inference. However, it has acquired a specific meaning synonymous with part of the `ignorability' assumption as stated by \citet{rosenbaum:1983}, that is, a certain conditional independence relationship between exposure (or treatment), potential outcomes, and possible confounding variables. In this paper, we study the links between the two usages of the term and point out their common underlying probabilistic arguments.  Furthermore, we propose a fully Bayesian formulation of causal inference that is based on exchangeable representations and includes Bayesian definitions of causal estimands.  Our central thesis is that de Finetti's formulation of exchangeability is entirely sufficient to give a coherent basis for causal inference, without the need to introduce special constructs (such as potential outcomes), mathematical machinery (such as the \textit{do}-operator), or additional conditional independence assumptions.

\subsection{Review of the de Finetti Representation for Exchangeable Sequences and the Problem Setup}

The de Finetti representation theorem for exchangeable sequences is a key mathematical result which underpins all Bayesian inference methodology. The original version for binary sequences was generalized to any real-valued random quantities by \citet{hewitt1955symmetric}, and the generalized version has been restated for example as Proposition 4.3 of \citet{bernardo:1994}. This states that if $(Y_i)_{i=1}^{\infty}$ is an infinite exchangeable sequence of random variables with probability law $\Pr$, there exists a random probability measure, $P$, such that conditionally on $P$, the $Y_n$ are independent and identically distributed (i.i.d.) with common distribution $P$. Moreover, with probability one, such a $P$ is the weak limit of the sequence of empirical distributions $\hat P_n(B) = \frac{1}{n} \sum_{i=1}^n \mathbf{1}_{\{Y_i \in B\}}$, where $B \subseteq \mathbb R$. In Bayesian learning under exchangeability, the random probability measure $P$ can be interpreted as an infinite-dimensional ``parameter'', with probability law, say $Q$, interpreted as the ``prior'' belief distribution. Hierarchically, this means that $Y_i \mid P \sim_{\textrm{i.i.d.}} P$ and $P \sim Q$. In the notation that follows we distinguish between the ``marginal'' measure, $\Pr$, and the random ``parameter-conditional'' measure, $P$, as the latter's existence is implied by the exchangeability property on the marginal distribution.

To characterize the conditional probability structures that appear in causal settings, we need the notion of \emph{partial exchangeability} originally introduced by \citet{definetti1938partial} and reviewed for example by \citet{diaconis1988recent}. Partial exchangeability characterizes the comparability of units within subpopulations that are formed, for example, by a (categorical) covariate. In our most basic setting, we have $W_i = (Y_i, Z_i, X_i)$ where $Y_i$ is an observable outcome, $Z_i$ is an observable treatment/exposure, and $X_i$ represents (typically a vector of) possible confounding variables. For simplicity, we consider the case where all variables take a finite number of possible values, possibly after discretizing continuous variables, so that $Y_i \in \mathcal Y \equiv \{0, 1, \ldots, \ell\}$, $Z_i \in \mathcal Z \equiv \{0, 1, \ldots, m\}$ and $X_i \in \mathcal X \equiv \{0, 1, \ldots, p\}$. However, we note that it is straightforward to generalize everything that follows to continuous outcomes $Y_i$ (see for example Definition 4.14 of \citealp{bernardo:1994}, for a generalization based on unrestricted exchangeability for sequences with predictive sufficient statistics).

For the joint distribution, for any $n \geq 1$ and combination of values $z_i \in \mathcal Z, x_i \in \mathcal X$ with a positive probability, we have the factorization
\begin{align*}
\MoveEqLeft[1] \Pr \left( \bigcap_{i=1}^n (Y_i = y_i, Z_i = z_i, X_i = x_i) \right) \\
&= \Pr \left( \bigcap_{i=1}^n (Y_i = y_i) \bigg | \bigcap_{i=1}^n (Z_i = z_i, X_i = x_i) \right) \\
&\quad\times \Pr \left( \bigcap_{i=1}^n (Z_i = z_i) \bigg | \bigcap_{i=1}^n (X_i = x_i ) \right) \\
&\quad\times \Pr \left( \bigcap_{i=1}^n (X_i = x_i)\right).
\end{align*}
Assuming exchangeability of the random vectors $W_i$ over the individual indices $i$, identity \eqref{eq:ex} becomes
\begin{align*}
\MoveEqLeft[1] \Pr \left( \bigcap_{i=1}^n (Y_i = y_i, Z_i = z_i, X_i = x_i) \right) \\
&= \Pr \left( \bigcap_{i=1}^n (Y_i = y_{\rho(i)}, Z_i = z_{\rho(i)}, X_i = x_{\rho(i)}) \right),
\end{align*}
where $\rho$ permutes the individual indices. By considering permutations $\rho$ that preserve the values of $Z$ and $X$ (so that $z_{\rho(i)} = z_i$ and $x_{\rho(i)} = x_i$), the exchangeability over $i$ also implies that 
\begin{align}\label{eq:deFpartialex}
\MoveEqLeft[1] \Pr \left(\bigcap_{i=1}^n (Y_i = y_i) \bigg | \bigcap_{i=1}^n (Z_i = z_i, X_i = x_i) \right) \\
&= \Pr \left( \bigcap_{z, x} \bigcap_{i \in I_{zx}^n} (Y_i = y_{\rho_{zx}(i)}) \bigg | \bigcap_{i=1}^n (Z_i = z_i, X_i = x_i) \right),\nonumber 
\end{align}
where $(z, x) \in \mathcal Z \times \mathcal X$ and $\rho_{zx}$ permutes the indices within the index set $I_{zx}^n = \{1, \ldots, n\} \cap \{i : Z_i = z, X_i = x\}$. Identity \eqref{eq:deFpartialex} corresponds to de Finetti's definition of partial exchangeability, and for example in the case of $\ell = 1$, implies the joint representation
\begin{align}\label{eq:deFpartial}
\MoveEqLeft[1] \Pr \left(\bigcap_{i=1}^n (Y_i = y_i) \bigg | \bigcap_{i=1}^n (Z_i = z_i, X_i = x_i) \right) \\
&= \int_{\mathcal P} \prod_{z, x} \prod_{i \in I_{zx}^n} P(Y_i = y_i \mid Z_i = z, X_i = x; \phi_{zx}) \,\textrm d Q(\phi), \nonumber 
\end{align}
where $\phi = (\phi_{00}, \ldots, \phi_{m p})$, $P(Y_i = y_i \mid Z_i = z, X_i = x; \phi_{zx}) = \phi_{zx}^{y_i}(1-\phi_{zx})^{1-y_i}$,
\begin{align*}
\MoveEqLeft[0.5] Q(\phi) \\
&= \lim_{n \longrightarrow \infty} \Pr\left\{ \bigcap_{z, x} \left(\frac{\sum_{i=1}^n \mathbf{1}_{\{Y_i = 1, Z_i = z, X_i = x\}}}{\sum_{i=1}^n \mathbf{1}_{\{Z_i = z, X_i = x\}}}  \le \phi_{zx} \right) \right\},
\end{align*}
and
\begin{equation*}
\phi_{zx} = \lim_{n \longrightarrow \infty} \frac{\sum_{i=1}^n \mathbf{1}_{\{Y_i = 1, Z_i = z, X_i = x\}}}{\sum_{i=1}^n \mathbf{1}_{\{Z_i = z, X_i = x\}}}.
\end{equation*}
The interpretation of \eqref{eq:deFpartial} is that within each treatment/covariate stratum the outcomes are conditionally independent and distributed as $Y_i \mid (Z_i = z, X_i = x; \phi_{zx}) \sim \textrm{Bernoulli}(\phi_{zx})$, and $Q$, which is a multivariate cumulative distribution function, is the prior belief distribution on the long-run, stratum-specific relative frequencies. Another interpretation is that the stratum-specific event counts are sufficient statistics with binomial distributions. The model specification would be completed by the specification of $Q$; a full discussion of the prior specification is beyond our scope here, but we note two special cases. Assuming $\phi_{00} = \ldots = \phi_{mp}$ would imply the exchangeability of the entire sequence (no difference between the groups), whereas assuming the group-specific parameters $\phi_{zx}$s themselves to be exchangeable would imply a hierarchical form for the representation (see for example Section 4.6.5 of \citealp{bernardo:1994}). We note that the latter property is different from the between-group exchangeability that we introduce in Section \ref{section:nounmeasuredconfounders} for causal considerations.

While representations such as \eqref{eq:deFpartial} enable statistical inferences on the unobservable characteristics of the infinite sequences based on observable finite sequences, further assumptions are needed for causal interpretations. Consider for example the case of $m=1$, with $Z_1 = 1$ and $Z_1 = 0$ representing the intervention and control groups, respectively. Here, the covariate stratum-specific risk differences $\phi_{1x} - \phi_{0x}$ or risk ratios $\phi_{1x}/\phi_{0x}$, or their marginal counterparts based on standardized risks $\sum_x \phi_{zx} P(X_i = x)$, would not be causal contrasts without further assumptions on the treatment assignment mechanism. We will argue that ruling out  unmeasured confounding requires a specific kind of between-group exchangeability in addition to the within-group property stated in \eqref{eq:deFpartialex}.

\subsection{Literature Review: Exchangeability and Causal Inference}

A connection between the original probabilistic concept of exchangeability and causal inference was first suggested by \citet[][p. 51]{lindley:1981}; however, the authors did not pursue this further. This connection was pointed out later by \citet{greenland:1986} in the context of non-identifiability of causal parameters due to confounding. However, in the causal inference literature \citep[e.g.,][]{greenland:1999,hernan:2006,greenland:2009}, `exchangeability' has been interpreted in terms of potential outcomes (instead of observable quantities), and the connection of this concept to its Bayesian interpretation appears to have been lost. In this paper, we highlight the similarities between causal reasoning based on unit-level exchangeability and the now more common formulation based on potential outcomes.

We aim to provide a sequel to the classic account of \citet{lindley:1981} that takes into account the numerous developments that have taken place in causal inference theory and methodology since. The utility of the concept of exchangeability and the account of \citet{lindley:1981} have been disputed by \citet[][p. 177--180]{pearl:2009(1)} \citep[see also][]{lindley:2002}, who argued that probability theory alone is not adequate for providing a comprehensive framework for causal reasoning (which, in fact, Lindley and Novick never attempted). Rather than enter this debate, we concentrate on clarifying the connection between the probabilistic notion of exchangeability and causal inference, using exchangeability as the basis of the `causal model.' A causal model is necessary to define the causal contrast of interest, as well as to define the notion of confounding and to state the identifying conditions required
for unconfounded inferences.

We follow the key insight of \citet[][p. 45]{lindley:1981} that ``inference is a process whereby one passes from data on a set of units to statements
about a further unit.'' Because we can only ever observe outcomes for any individual unit under a single exposure pattern, it seems reasonable to base
statistical inferences about causal effects on an explicit assumption of `similarity' (or more precisely, indistinguishability) of the individual
instances. To assume an exchangeable structure is always appropriate after sufficient relevant information has been included \citep[][p. 6]{gelman:2004};
however, what constitutes sufficient relevant information in causal inference settings often has to be decided based on prior information alone, as
noted by \citet{greenland:2009}. That is, causal inferences from observational settings necessarily rely on prior information regarding the causal
mechanisms involved; the role of prior information can be made explicit in Bayesian inference.

Several other authors have attempted to make connections between classical statistical models and causal models. In particular
\citet{dawid:2000}, \citet{arjas:2004}, and \citet{chib:2007} have suggested that the potential outcomes notation is redundant in formulating causal models, and similar arguments have been made both in Bayesian and frequentist settings. \citet{baker2013causal} gave a probabilistic interpretation to confounding and collider biases. Many of the formulations put forth as alternatives to potential outcomes are based on introducing a hypothetical `randomized' or `experimental' probability measure that is used to formulate the causal quantity of interest \citep{dawid:2010,roysland:2011,arjas:2012,saarela:2015,commenges2019causality}. Inference then becomes a matter of linking the experimental measure to the observational one thought to have generated the data, which involves assumptions about the absence of unmeasured confounding. Other formulations are based on structural definitions, where a deterministic relationship is assumed between observed and latent variables \citep{commenges2015stochastic,ferreira2019causality}.

The `no confounding' assumption required for identification of the causal effect under these formulations is usually expressed in terms of latent variables, or equivalence of certain components of the experimental and observational joint distributions, termed by \citet{dawid:2010} as the `stability' assumption. \citet{buhlmann2020invariance} termed a similar property `invariance' and formulated causal inference in terms of a risk minimization problem. \citet{ferreira2015some} framed an exchangeability property concerning the treatment assignment mechanism as a `no confounding' type assumption, but they did not connect it to Bayesian inference. We are not aware of exchangeability (in its original meaning as a symmetry property of probability distributions) otherwise used as a causal assumption; \citet{dawid2016statistical} used it as an inferential assumption needed in addition to a `no confounding' type assumption. \citet{dawid2021decision} made a distinction between post-treatment and pre-treatment exchangeability, where the latter is closely related to the notion of partial exchangeability of outcomes within treatment and control groups separately, while the former involves a judgment of similarity of the groups being compared before they received treatment. A further ignorability condition concerning the treatment assignment mechanism is needed for causal inferences based on the observed responses in the treatment and control groups. 

Like \citet{dawid2021decision}, we consider partial exchangeability, as defined above, as a starting point, suggesting parametric inferences based on the representation theorem. However, while this within-group exchangeability is sufficient for predicting the outcome for a further exchangeable unit, causal inferences require a judgment on exchangeability \emph{between} groups, that is, between treated and untreated units, reflecting the absence of confounding due to the group characteristics. In this work, our primary objective is to formulate the required condition as a probabilistic symmetry property. Furthermore, we show that this property indeed is an identifiability condition for causal effects as it implies ignorability of the treatment assignment mechanism. Under this condition, the parameters suggested by the representation theorem have a causal interpretation, which provides a link to Bayesian causal inferences. We further extend this reasoning to longitudinal settings, where in addition to biases due to confounding, we can encounter biases related to conditioning on intermediate variables. Similar to \citet{ferreira2019causality}, we adopt a structural model notation as this allows us to draw connections between the different causal models but with a focus on Bayesian causal inference.

\subsection{Manuscript Outline}

The paper proceeds as follows. In Section \ref{section:foundations}, we introduce the necessary notation and concepts. In Section \ref{section:nounmeasuredconfounders}, we propose a definition of conditional exchangeability to be used as an identifying condition for estimating causal effects. We show that this condition implies ignorability of the treatment assignment mechanism and relate it to alternative conditions based on causal diagrams and potential outcomes. In Section \ref{section:causalcontrasts}, we give a Bayesian definition of a marginal causal contrast and consider inference under observational settings. In Section \ref{section:nullparadox}, we consider extending the proposed framework to longitudinal settings. We conclude with a discussion in Section \ref{section:discussion}.

\section{Notation and foundations}\label{section:foundations}

\subsection{Structural Assumption}\label{section:structural}

It is convenient for our derivations to assume that the outcome random variable, $Y_i$, is determined by the structural rule $Y_{i} = f(Z_i,X_i,U_i)$,
where $Z_i$ represents treatment assignment, $X_i$ observed potential confounders, and $U_i$ unobserved 
factors that may be determinants of $Y_i$ and may or may not also be confounders. This structural assumption is quite general as the model can be readily modified to include further stochastic elements such as additive `residual' errors. Note that in the structural definition, we may consider specific \textit{interventions} on treatment and write
$f(z,X_i,U_i)$, as if random variable $Z_i$ has a degenerate distribution at $z$, and so that the intervention is independent of $(X_i,U_i)$. Note
also that the structural definition is essentially identical to the potential outcome construction; in the conventional notation, the
potential outcome is given by $Y_i(z) \equiv f(z,X_i,U_i)$. In what follows, we always assume `general' infinite exchangeability of the sequence $\left( (Z_i, X_i, U_i) \right)_{i=1}^\infty$ (and consequently $(Y_i)_{i=1}^\infty$ as it is determined by the former) over the individual indices $i$, which also implies exchangeability of the sequence $(W_i)_{i=1}^\infty$ of the observable random vectors $W_i = (Y_i, Z_i, X_i)$. For finite sequences of these, in places we use vector notation $(W_i)_{i=1}^n = (W_1, \ldots, W_n)$.

\subsection{Experimental and Observational Designs}

The objective of causal inference is to quantify the effect of assigning a treatment level, $z$, (relative to an alternative level $z'$) on the outcome, independent of any other determinants of the outcome. Such an allocation mechanism is commonly termed \textit{experimental}. We label the corresponding probability distributions of observations under such a setting by $\mathcal E$. If the independence is not known to be present, the mechanism is termed \textit{observational}, or \emph{non-experimental}. The corresponding distributions are labeled by $\mathcal O$. The independence requirement may be expressed as the factorization
\begin{align}\label{eq: Expfactorize}
\MoveEqLeft[1] \Pr \left( \bigcap_{i=1}^n (Z_i = z_i, X_i = x_i, U_i \in \textrm du_i; \mathcal E \right) \\
&=  \Pr \left( \bigcap_{i=1}^n (Z_i = z_i)  ; \mathcal E \right) \nonumber \\
&\quad\times \Pr \left( \bigcap_{i=1}^n (X_i = x_i, U_i \in \textrm du_i) ; \mathcal E \right), \nonumber
\end{align}
for any $n \geq 1$, where each of the factors on the right-hand side has a representation of the form of \eqref{eq:deFpartial}. From this it also follows that $Z_j \independent (X_k, U_k)$ for all $j, k$, where we use $\independent$ to denote statistical independence. This expression could be generalized to allow the treatment assignment to depend on the observed characteristics $X_i$, but in what follows we proceed with \eqref{eq: Expfactorize}.

\subsection{Directed Acyclic Graphs}

In subsequent sections, our explanations are assisted by the use of directed acyclic graphs (DAGs) to illustrate the underlying relationships between the variables. In the Bayesian framework, we can regard a posited DAG as encapsulating structural prior knowledge related to the observable quantities, and they may be considered either conditional on or marginalized over parameters in models. In this paper we use the terms `knowledge,' `information,' and `opinion' interchangeably to describe the \textit{a priori}-held subjective beliefs -- both qualitative and quantitative -- of the experimenter. As a notational device, we will use structural definitions to illustrate the link between the information encoded in a DAG and the corresponding probability statements.

The DAG in the left-hand panel of Figure \ref{figure:dag1randomized} illustrates the relationship between variables as described in Section \ref{section:structural} and where $Z$ is assigned experimentally. Figure \ref{figure:dag1randomized} relates to a single individual $i$; by convention, in the frequentist setting, the nodes on a single DAG are interpreted to indicate probabilistic relationships for random variables relating to an archetypal individual present in a random sample, with the graph replicated identically across the independent draws $i=1,\ldots,n$. As indicated by equation \eqref{eq:deFpartial}, however, under an assumption of exchangeability, the collections of variables $W_i, i=1,\ldots,n$ are not marginally independent, but instead are conditionally independent given parameter $P$.  Under the assumption of exchangeability of the $W_i$, the most general DAG would have an additional node containing $P$ from which arrows into the complete collection of variables would emanate (Figure A1 in Supplementary Appendix A; \citealp{supplement}).

\tikzset{
    -Latex,auto,node distance =1 cm and 1 cm,semithick,
    state/.style ={circle, draw, minimum width = 0.7 cm},
    box/.style ={rectangle, draw, minimum width = 0.7 cm, fill=lightgray},
    point/.style = {circle, draw, inner sep=0.08cm,fill,node contents={}},
    bidirected/.style={Latex-Latex,dashed},
    el/.style = {inner sep=3pt, align=left, sloped}
}
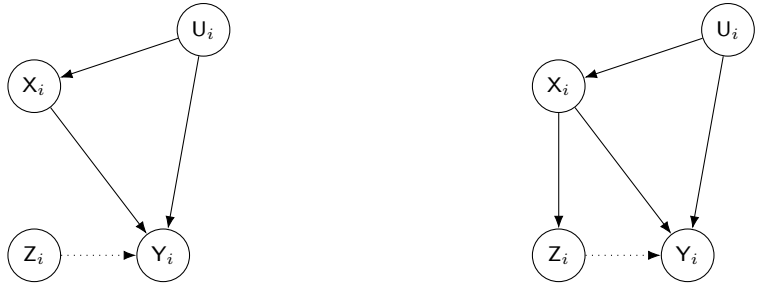
\begin{figure}[!hb]
\centering
\begin{tikzpicture}[scale=1.5]
    \node[state] (z) at (0,0) {$\textsf{Z}_i$};
    \node[state] (x) at (0,1.5) {$\textsf{X}_i$};
    \node[state] (u) at (1.5,2) {$\textsf{U}_i$};

    \node (e) at (-0.5,2) {$\mathcal E$};

    \node[state] (y) [right =of z] {$\textsf{Y}_i$};

    \path (z) edge[dotted] (y);
    \path (x) edge (y);
    \path (u) edge (y);
    \path (u) edge (x);


\end{tikzpicture}
\hspace{0.25in}
\begin{tikzpicture}[scale=1.5]
    \node[state] (z) at (0,0) {$\textsf{Z}_i$};
    \node[state] (x) at (0,1.5) {$\textsf{X}_i$};
    \node[state] (u) at (1.5,2) {$\textsf{U}_i$};

    \node (o) at (-0.5,2) {$\mathcal O$};

    \node[state] (y) [right =of z] {$\textsf{Y}_i$};

    \path (z) edge[dotted] (y);
    \path (x) edge (z);
    \path (x) edge (y);
    \path (u) edge (y);

    \path (u) edge[solid] (x);

\end{tikzpicture}
\caption{Left-hand panel: DAG depicting a randomized setting labeled $\mathcal E$.  The dotted arrow $Z_i \longrightarrow Y_i$ is absent if there is no treatment effect. Right-hand panel: DAG depicting an observational setting labeled $\mathcal O$.  The arrow $X_i \longrightarrow Z_i$ is the distinguishing feature of $\mathcal O$ compared to $\mathcal E$; conditioning on $Z_i$ would open a confounding `backdoor' path from $Z_i$ to $Y_i$.}
\label{figure:dag1randomized}
\end{figure}

\section{Exchangeability and ignorability}\label{section:nounmeasuredconfounders}

\subsection{Exchangeability Under Randomization}
\label{sec:Ex_ind}

Under the randomized setting $\mathcal E$, factorization \eqref{eq: Expfactorize} and the general exchangeability of $(X_i, U_i)$ imply an additional exchangeability property that we can give a causal interpretation. A similar property can then be considered as an identifying assumption for causal effects in an observational setting $\mathcal O$, where this property is not implied by design. Now, taking $A \equiv \{Z_{j} = z, Z_{k} = z'$\} to be the observed treatment assignment,
\begin{align}
\label{equation:populationexchangeabilitye}
\MoveEqLeft[1] \Pr(Y_{j} = y, Y_{k} = y' \mid A; \mathcal E) \\
&= \Pr\big(f(z,X_j,U_j) = y, f(z',X_k,U_k) = y' \mid A; \mathcal E\big) \nonumber \\
&= \Pr\big(f(z,X_j,U_j) = y, f(z',X_k,U_k) = y' ; \mathcal E\big) \nonumber \\
&= \Pr\big(f(z,X_k,U_k) = y, f(z',X_j,U_j) = y' ; \mathcal E\big) \nonumber \\
&= \Pr\big(f(z,X_k,U_k) = y, f(z',X_j,U_j) = y' \mid A; \mathcal E\big) \nonumber
\end{align}
for all $(y,y')$ and $(z,z')$. Here the first equality followed from the functional definition, third from exchangeability and second and fourth from independence of the assignment mechanism. In particular, \eqref{equation:populationexchangeabilitye} states that under the experimental setting, the joint distribution of the two outcomes is the same under a hypothetical switch of the interventions. Thus, taking $z = 1$ and $z' = 0$ and $A \equiv \{Z_{j} = 1, Z_{k} = 0$\}, the property
\begin{align*}
\MoveEqLeft \Pr\big(f(1,X_j,U_j) = y, f(0,X_k,U_k) = y' \mid A; \mathcal E\big) \nonumber \\
&= \Pr\big(f(1,X_k,U_k) = y, f(0,X_j,U_j) = y' \mid A; \mathcal E\big) \nonumber
\end{align*}
suggests a causal interpretation; the joint distribution of the outcomes does not depend on which individual was actually assigned treatment $z = 1$. In other words, the known treatment assignment $A$ is not informative of the other determinants of the outcomes. This property does not follow from the previously assumed exchangeability over $i$
\begin{align*}
\MoveEqLeft[1] \Pr(Y_{j} = y, Y_{k} = y' \mid Z_{j} = 1, Z_{k} = 0; \mathcal E) \\
&= \Pr(Y_{k} = y, Y_{j} = y' \mid Z_{j} = 0, Z_{k} = 1; \mathcal E),
\end{align*}
that is, even under the experimental setting, the statement
\begin{align*}
\MoveEqLeft[1] \Pr(Y_{j} = y, Y_{k} = y' \mid Z_{j} = 1, Z_{k} = 0; \mathcal E) \\
&= \Pr(Y_{k} = y, Y_{j} = y' \mid Z_{j} = 1, Z_{k} = 0; \mathcal E)
\end{align*}
would only be true if there were no treatment effect. While we could consider such `under the null' causal exchangeability statements, the structural model
allows us to make explicit the hypothetical switching of the treatments without this restriction.

Statement \eqref{equation:populationexchangeabilitye} can be extended to any finite sequence of observations, conditional on a sequence of treatment
assignments, as
\begin{align}\label{exinf}
\MoveEqLeft[1] \Pr \left(\bigcap_{i=1}^n  (f(z_i,X_i,U_i) = y_{i}) \bigg | \bigcap_{i=1}^n (Z_{i} = z_i); \mathcal E  \right) \\
&= \Pr \left(\bigcap_{i=1}^n (f(z_i,X_{\rho(i)},U_{\rho(i)}) = y_{i}) \bigg | \bigcap_{i=1}^n (Z_{i} = z_i); \mathcal E  \right) \nonumber 
\end{align}
for any permutation $\rho(.)$ of the indices. In the remainder of this section we show that under property \eqref{exinf}, the parameters implied by representation \eqref{eq:deFpartial}, such as contrasts of treatment group specific outcome frequencies/risks, have a causal interpretation. We note first that by \eqref{exinf}, considering permutations only within the treatment groups, the sequences of `treated' random variables $f(1,X_{i},U_{i})$ and `untreated' random variables $f(0,X_{i},U_{i})$ are partially exchangeable. Thus, the within-group exchangeability of outcome sequences \eqref{eq:deFpartialex} is a special case of \eqref{exinf}, the interpretation being that the stronger condition extends partial exchangeability to certain kinds of between-group comparisons. Essentially, \eqref{exinf} states that the remaining determinants, observed and unobserved, of the outcomes are exchangeable between the treatment groups. We note that \eqref{exinf} would follow from assuming $X_i$ and $U_i$ to be similarly exchangeable, but this would be an unnecessarily strong assumption, as in \eqref{exinf} this is only required for the aspects of $X_i$ and $U_i$ that are determinants of the outcome.

We term property \eqref{equation:populationexchangeabilitye} and its extension \eqref{exinf} as `conditional exchangeability' to distinguish them
from the previously assumed partial exchangeability. While under the experimental setting these were implied by the latter and the known properties of
the treatment assignment mechanism, under observational settings considered in Section \ref{section:individuallevel}, a similar property will have to be
assumed \textit{a priori}. When the assignment mechanism is unknown, this is a strong assumption, but one that is needed for the identifiability of causal effects based on observational studies. It then becomes important that the conditional exchangeability statements imply properties of the treatment assignment mechanism. To see this, from
\eqref{equation:populationexchangeabilitye} it follows that
\begin{align*}
\MoveEqLeft[1] \textstyle \sum_{y'} \Pr\big(f(1,X_j,U_j) = y, f(0,X_k,U_k) = y' \mid A; \mathcal E\big) \nonumber \\
&= \textstyle \sum_{y'} \Pr\big(f(1,X_k,U_k) = y, f(0,X_j,U_j) = y' \mid A; \mathcal E\big) \nonumber \\
\Rightarrow& \Pr\big(f(1,X_j,U_j) = y \mid A; \mathcal E\big) \nonumber \\
&\quad= \Pr\big(f(1,X_k,U_k) = y \mid A; \mathcal E\big),
\end{align*}
that is, $f(1,X_j,U_j) \mid  (Z_{j} = 1, Z_{k} = 0) \stackrel{\textrm{d}}{=} f(1,X_k,U_k) \mid  (Z_{j} = 1, Z_{k} = 0)$ under $\mathcal E$. If we further
assume that the treatment assignment of individual $k$ is not informative of the outcome of individual $j$ and vice versa (corresponding to the common
assumption of `no interference between units', cf. \citealp{rubin:1978}; \citealp[][p. 58]{lindley:1981}) we have that \[f(1,X_j,U_j) \mid  (Z_{j} = 1) \stackrel{\textrm{d}}{=} f(1,X_k,U_k) \mid  (Z_{k} = 0).\] Further, by general exchangeability we have that
\[f(1,X_j,U_j) \mid  (Z_{j} = 1) \stackrel{\textrm{d}}{=} f(1,X_k,U_k) \mid (Z_{k} = 1),\] and combining this with the previous, that $f(1,X_i,U_i)
\independent Z_{i}$ under $\mathcal E$. By a symmetrical argument, we can show that $f(0,X_i,U_i) \independent Z_{i}$, and finally that $f(z,X_i,U_i) \independent Z_{i}$, $z \in \{0,1\}$ in the $\Pr$ distribution. This independence property was implied by \eqref{equation:populationexchangeabilitye} and the no interference between units assumption. 

While the previous applies marginally, symmetry property \eqref{exinf} holds true also conditional on the parameters implied by the de Finetti representation, following the arguments in the Appendix. We also note that in the $P$ distribution, the `no interference between units' property is implied by the general exchangeability due to the resulting i.i.d. structure. Thus, we also have $f(z,X_i,U_i) \independent Z_{i}$ in the $P$ distribution, which is equivalent to the ignorability condition $Y_i(z) \independent Z_i$ commonly stated in terms of potential outcomes. We return to this connection in Section \ref{section:potential} but note that under the randomized setting, we have demonstrated that exchangeability and ignorability both express a similar `no confounding' property. This property allows for unconfounded comparisons of the treatment arms in terms of long-run outcome frequencies. Expressing this as a probabilistic symmetry statement allows us to make use of Bayesian concepts in outlining a causal modeling framework. A perceived strength of the potential outcomes framework is being able to express causal contrasts of interest directly in terms of the average potential outcomes, such as $E[Y(1)] - E[Y(0)]$ and $E[Y(1)]/E[Y(0)]$ for risk difference and ratio, respectively, without referring to parameters in statistical models. Similar constructs are also possible in the present framework, which we will address in Section \ref{section:causalcontrasts}.

\subsection{Exchangeability in the Observational Setting} \label{section:individuallevel}

While it was helpful to demonstrate the ideal properties of the experimental setting, we are actually interested in inferences under observational settings, where we do not choose the treatment assignment mechanism and the exchangeability of the subpopulations being compared does not follow from the study design. We consider a hypothetical study of the effect of initiation of antiretroviral therapy on CD4 cell counts, based on a cohort of $n$ HIV patients. Specifically, for $i=1,\ldots,n$, let random variable $X_i$ represent a baseline CD4 cell count measurement for HIV-positive individual $i$, $Z_i$ represent the decision to initiate antiretroviral therapy at the baseline time point, and $Y_i$ the CD4 cell count measurement after a fixed time has passed since  baseline. Further, let $U_i$ be a latent variable representing the underlying immune status of individual $i$, some facet of which could possibly be captured by $X_i$. We know that individuals with lower CD4 cell counts, $X_i$, are more likely to initiate treatment; the correlation between $U_i$ and $X_i$ further implies that those with weakened underlying immune function are more likely to initiate treatment. Thus, the factorization in \eqref{eq: Expfactorize} likely does not hold. Such dependencies are illustrated in the right-hand DAG in Figure \ref{figure:dag1randomized}.

If we are interested in the causal effect of treatment initiation, it seems appropriate to formulate the \emph{causal estimand} of interest in terms of a hypothetical randomized trial that otherwise resembles the observational setting but where \eqref{eq: Expfactorize} holds. The structural assumption $Y_{i} = f(Z_i,X_i,U_i)$ can be taken to apply under both settings, and the distribution of $(X_i,U_i)$ can also be assumed to be the same as we don't actually observe any data under $\mathcal E$. The problem of causal inference then involves making probability statements about the estimand specified under $\mathcal E$ based on data collected under $\mathcal O$. Based on the context, a simple comparison in terms of the summary statistics of the outcomes among those observed to be treated and those observed not would be \textit{confounded}; a numerical example demonstrating this is presented in Supplementary Appendix B. We formalize this concept in Section 4, where we introduce an explicit causal estimand and its estimator. Before that, we attempt to understand this non-comparability of the groups through a probabilistic exchangeability statement between individuals representative of those groups and propose this statement as an identifying condition for causal effects.

Consider a comparison of two individuals, $j$ and $k$, with observed treatment assignments $Z_j = 1$ and $Z_k = 0$, respectively, but with
the outcome yet to be observed. As in Section \ref{sec:Ex_ind}, we consider whether the outcomes of these two individuals are exchangeable
(pairwise) under a hypothetical intervention to reverse their treatments; if so, a comparison of their outcomes under the actual treatment assignments
would be informative of the causal effect of the treatment. We again take the outcome to be determined by a structural model, in which case the required exchangeability property conditional on $A \equiv \{Z_{j} = 1, Z_{k} = 0\}$ can be expressed as
\begin{align}\label{equation:populationexchangeability}
\MoveEqLeft[1] \Pr(Y_{j} = y, Y_{k} = y' \mid A; \mathcal O) \\
&= \Pr\big(f(1,X_j,U_j) = y, f(0,X_k,U_k) = y' \mid A; \mathcal O\big) \nonumber \\
&= \Pr\big(f(0,X_j,U_j) = y', f(1,X_k,U_k) = y \mid A; \mathcal O\big) \nonumber
\end{align}
for all $(y,y')$, which mirrors the property obtained under $\mathcal E$. We emphasize that a statement such as \eqref{equation:populationexchangeability} could usually only be made on a subjective basis, conditional on information concerning the study design and data generating mechanism; it represents a strong assumption requiring no
unmeasured confounding. In addition, the causal question of interest, including the role of the variables in the data generating mechanism, must be
stated \emph{a priori}; without this knowledge, we would not know which exchangeability judgment is relevant for drawing causal inferences.
Identity \eqref{equation:populationexchangeability} could be extended to any finite sequence similar to \eqref{exinf}. If this property holds under $\mathcal
O$, and we additionally assume no interference between units, we would obtain $Z_i \independent f(z,X_i, U_i)$
under $\mathcal O$.

Central to (\ref{equation:populationexchangeability}) for causal considerations is the extent to which group assignment can tell us about the other
characteristics of the groups through the \emph{a priori} knowledge of the relationships between the variables. If statement
(\ref{equation:populationexchangeability}) was true, the treatment and reference groups, and individuals $j$ and
$k$, would be directly comparable, implying that a comparison of the two groups through a suitable summary statistic, for instance
$$\frac{\sum_{i=1}^n z_i y_i}{\sum_{i=1}^n z_i} - \frac{\sum_{i=1}^n (1-z_i) y_i}{\sum_{i=1}^n (1-z_i)}$$
would be free from confounding. However, exchangeability of the units of inference implies that the \emph{labels of the units do not carry relevant
information} (\citealp[e.g.,][p. 168]{bernardo:1994}; \citealp[][p. 6]{gelman:2004}), which is now clearly not the case because of how the comparison was constructed: \emph{a priori} we
would expect individual $j$ to have lower baseline CD4 count than $k$ based on the treatment assignments.

\subsection{Restoring Exchangeability Through Conditioning.}\label{section:conditioning}

The strong assumption in \eqref{equation:populationexchangeability} can be weakened using conditioning. In the example above, if the baseline CD4 count sufficiently represents the indication to initiate treatment, we can stratify on this variable to achieve better comparability. Let now $j$ and $k$ index treated ($Z_j = 1$) and untreated ($Z_k = 0$) individuals matched on the condition $X_j=X_k=x$ (say, baseline CD4 count). Now, we could assume conditional on $A_x \equiv \{Z_{j} = 1, Z_{k} = 0, X_{j} = X_{k} = x\}$ that
\begin{align}\label{equation:spope}
\MoveEqLeft[1] \Pr(Y_{j} = y, Y_{k} = y' \mid A_x; \mathcal O) \\
&= \Pr\big(f(1,x,U_j) = y, f(0,x,U_k) = y' \mid A_x; \mathcal O\big) \nonumber \\
&= \Pr\big(f(0,x,U_j) = y', f(1,x,U_k) = y \mid A_x; \mathcal O\big) \nonumber
\end{align}
for all $(y,y')$. Similar to the discussion in Section \ref{sec:Ex_ind}, \eqref{equation:spope} implies that $f(z,x,U_j) \mid  (Z_{j} = 1,
Z_{k} = 0, X_{j} = X_{k} = x) \stackrel{\textrm{d}}{=} f(z,x,U_k) \mid  (Z_{j} = 1, Z_{k} = 0, X_{j} = X_{k} = x)$ under $\mathcal O$. And further, under the assumption of no interference  between the units, $Z_i \independent f(z, x, U_i) \mid X_i = x$ under $\mathcal O$. Condition \eqref{equation:spope} can be
extended to any finite sequence matched on $x$, similar to \eqref{exinf}. Because \eqref{equation:spope} applies also under the experimental setting $\mathcal E$
and we assume the distribution of the baseline characteristics $(X_i,U_i)$ to be the same in both $\mathcal O$ and $\mathcal E$, we also have that
\begin{align*}
\MoveEqLeft[1] \Pr(Y_i \mid Z_i = z, X_i = x; \mathcal E) \\
&= \Pr(f(z,x,U_i) \mid Z_i = z, X_i = x; \mathcal E) \\
&= \Pr(f(z,x,U_i) \mid X_i = x; \mathcal E) \\
&= \Pr(f(z,x,U_i) \mid X_i= x; \mathcal O) \\
&= \Pr(f(z,x,U_i) \mid Z_i = z, X_i = x; \mathcal O) \\
&= \Pr(Y_i \mid Z_i = z, X_i = x; \mathcal O).
\end{align*}
Following the arguments in the Appendix, the same property would also apply in the i.i.d. distribution implied by the infinite exchangeability. The corresponding equivalence
\[
P(Y_i \mid Z_i, X_i; \mathcal E) = P(Y_i \mid Z_i, X_i; \mathcal O)
\]
is the `no confounding' condition termed \textit{stability} by \citet{dawid:2010}. Here exchangeability implies stability for the conditional
outcome distribution, which is one of the required identifying conditions for inferences on marginal causal contrasts (Section
\ref{section:estimation}).

\subsection{Connection to Posterior Predictive Inferences}\label{section:ppred}

Comparison \eqref{equation:spope} involved two individuals with an opposite treatment assignment and is the relevant comparison for causal considerations.
For predictive considerations, the general infinite exchangeability is sufficient, implying the partial exchangeability of the outcomes within subgroups with the same characteristics, that is, 
\begin{align}\label{equation:spaire}
\MoveEqLeft[1] \Pr(Y_j = y, Y_k = y' \mid Z_j = Z_k = z, X_j = X_k = x; \mathcal O) \\
&= \Pr(Y_j = y', Y_k = y \mid Z_j = Z_k = z, X_j = X_k = x; \mathcal O). \nonumber
\end{align}
The pairwise exchangeability statement \eqref{equation:spaire} extends from observed units $i = 1, \ldots, n$ to further similarly matched units $j$ and $k$, $j, k > n$, which motivates the use of posterior predictive inferences within the treatment groups. The further conditional exchangeability consideration \eqref{equation:spope} suggests that the predictions can be compared across the treatment groups as
\newlength{\templen}
\settowidth{\templen}{$\approx\sum_{x} \bigg($}
\begin{align}\label{equation:ppdirectstandardization}
\MoveEqLeft[1] \sum_{x}  \big( E[Y_{j} \mid Z_{j} = 1, X_{j} = x, D_{1x}; \mathcal O] \\
&\quad - E[Y_{k} \mid Z_{k} = 0, X_{k} = x, D_{0x}; \mathcal O] \big) \frac{1}{n} \sum_{i=1}^n \mathbf 1_{\{X_i = x\}} \nonumber \\
&\approx\sum_{x} \bigg(\frac{\sum_{i=1}^n \mathbf 1_{\{X_i = x\}} z_i y_i}{\sum_{i=1}^n \mathbf 1_{\{X_i = x\}} z_i} \nonumber \\
&\hspace*{\templen}- \frac{\sum_{i=1}^n \mathbf 1_{\{X_i = x\}} (1-z_i) y_i}{\sum_{i=1}^n \mathbf 1_{\{X_i = x\}} (1-z_i)} \bigg) \frac{1}{n} \sum_{i=1}^n \mathbf 1_{\{X_i = x\}}, \nonumber 
\end{align}
where $D_{zx} \equiv \{W_i : i \in I_{zx}^n\}$ denotes the observed data on the matched groups and where the last form follows by approximating the within-stratum posterior predictive means by the sample means \citep[cf.][p. 47]{lindley:1981}. Thus, the above recovers the classical direct standardization formula for the marginal treatment effect (\citealp{keiding2014standardization}, also known as the back-door adjustment formula, \citealp{pearl:2009(1)}).

For Bayesian inference, if the strata are too small for the use of the direct standardization formula, one would instead have to pool the observed data and
connect them to the predictions through parametric probability models. We will formalize this in the following section but note here that the modeling
approach requires the existence of parameter vectors $\Phi$ and $\Psi$ given which $Y_{j} \independent (W_i)_{i=1}^n \mid (Z_{j}, X_{j}, \phi)$ and
$X_{j} \independent (X_i)_{i=1}^n \mid \psi$ under $\mathcal O$, for all $j = n+1,\ldots$. As outlined in Section \ref{section:introduction}, the existence of such parameters is implied by the partial or unrestricted exchangeability assumptions. Given an observed realization $(w_i)_{i=1}^n$, a parametric counterpart to (\ref{equation:ppdirectstandardization}) can be given as 
\settowidth{\templen}{$\displaystyle \int_{\phi, \psi} \sum_{x}\big($}
\newlength{\templennew}
\settowidth{\templennew}{$\displaystyle \int_{\phi, \psi} \sum_{x}$}
\begin{align}\label{equation:ppparametricdirectstandardization}
&\int_{\phi, \psi} \sum_{x} \big (E[Y_{j} \mid Z_{j} = 1, X_{j} = x, \phi; \mathcal O] \\
&\hspace*{\templen}-E[Y_{k} \mid Z_{k} = 0, X_{k} = x, \phi; \mathcal O] \big) \nonumber\\
&\hspace*{\templennew}\times P(X_j = x \mid \psi; \mathcal O) \,\textrm d Q(\phi, \psi \mid (w_i)_{i=1}^n; \mathcal O). \nonumber
\end{align}
It is apparent from \eqref{equation:ppparametricdirectstandardization} that drawing causal
inferences is possible if the stability property of Section \ref{section:conditioning} applies to the pairwise comparisons
\begin{align*}
\MoveEqLeft[1] E[Y_{j} \mid Z_{j} = 1, X_{j} = x, \phi; \mathcal O] \\
&- E[Y_{k} \mid Z_{k} = 0, X_{k} = x, \phi; \mathcal O],
\end{align*}
with $\phi$ parametrizing the causal effect of $Z_i$ on $Y_i$ when controlling for $X_i$ as in this case, the inferences
would be the same as under the experimental design. However, parametrizing causal effects directly would be reliant on statistical models, whereas
the convention in causal inference literature, especially in potential outcome formulations, is to define the causal contrasts of interest first without
reference to models. We address model-free definitions of causal contrasts in the present framework in Section \ref{section:definition}, where parametric
models may then be utilized to obtain estimators for such contrasts.

\subsection{Connection to Other Latent Variable Formulations}\label{section:bd}

Under the point treatment setting, the implications of the infinite extension of criterion \eqref{equation:spope} are equivalent to other conditions for unconfounded inferences stated in terms of conceptual latent variables representing general confounding. For instance, Definition 1 of \citet{arjas:2012} connects unconfounded inferences to the conditional independence property $Z_i \independent U_i \mid X_i$. This in turn directly implies that $Z_i \independent f(z, x, U_i) \mid X_i = x$, and further the stability property similarly to Section \ref{section:conditioning}.

Although formulations in terms of latent variables need not rely on causal graphs, the absence of unmeasured confounders can be stated equivalently in terms of the back-door criterion of \citet[][p. 79]{pearl:2009(1)}; in the absence of a direct arrow $U_i \longrightarrow Z_i$ in the right-hand panel of Figure \ref{figure:dag1randomized}, $X_i$ blocks every path between $Z_i$ and $Y_i$ that contains an arrow into $Z_i$ (and is not a descendant of $Z_i$), which implies that $X_i$ is sufficient to control for confounding. Alternatively, the conditional independence property $Z_i \independent U_i \mid X_i$ can be read directly from the graph of Figure \ref{figure:dag1randomized} using, for example, the moralization criterion of \citet{lauritzen1989mixed}.

\subsection{Connection to the Potential Outcomes Notation}\label{section:potential}

Under the structural definition of the outcome, we can take the potential outcomes of individual $i$ to be determined by $Y_{i}(z) = f(z,X_i,U_i)$ \citep[cf.][p. 98]{pearl:2009(1)}, with the observed outcome given by $Y_i = Y_{i}(Z_i)$ (the latter is known as the \emph{consistency} assumption, \citealp[e.g.,][]{cole:2009,vanderweele:2009,pearl:2010}). As discussed in the previous two sections, the infinite extension of the symmetry property \eqref{equation:spope} implies that $Z_i \independent f(z, x, U_i) \mid X_i = x$ under $\mathcal O$. This is equivalent to the statement
\begin{align}
\label{equation:conditionalexchangeability2}
Y_{i}(z) \independent Z_i \mid X_i,
\end{align}
which is in fact the probabilistic `conditional exchangeability' condition as defined by \citet[][p. 579]{hernan:2006}, or a consequence of the first part of the \emph{strongly ignorable treatment assignment} condition, as defined by \citet[][p. 43]{rosenbaum:1983}. We note that making statements about the joint distribution of the potential outcomes (as in strong ignorability) is not necessary for identification of causal contrasts; the weak version involving \eqref{equation:conditionalexchangeability2} suffices. Although \citet[][p. 41] {rubin:1978} uses the term exchangeability in the usual Bayesian sense to justify an i.i.d. model construction, as far as we know, the connection between the Bayesian notion of exchangeability and the condition stated in terms of potential outcomes has not been made or studied within the framework of Rubin's causal model \citep[as termed by][]{holland:1986}. In contrast, this connection is implied in \citet{greenland:1986}, \citet{greenland:1999} and \citet{greenland:2009}.

We note that, under the probabilistic exchangeability condition \eqref{equation:spope}, we had that $f(z,x,U_j) \mid  (Z_{j} = 1, Z_{k} = 0, X_{j} = X_{k} = x) \stackrel{\textrm{d}}{=} f(z,x,U_k) \mid  (Z_{j} = 1, Z_{k} = 0, X_{j} = X_{k} = x)$, that is, the remaining determinants of the outcome under the structural model have the same population distribution between the treatment groups. Requiring that these determinants also have the same empirical distribution between the groups being compared would correspond to the \textit{deterministic exchangeability} condition laid out by \citet[][p. 415]{greenland:1986}. This is unnecessarily strong for unconfounded inferences; it rules out both confounding and imbalance (e.g., the chance imbalances that could arise even under complete randomization). If we could condition on all of the determinants of the outcome, the symmetry property conditional on $A_{xu} \equiv \{Z_{j} = 1, Z_{k} = 0, X_{j} = X_{k} = x, U_{j} = U_{k} = u\}$ could be written as
\begin{align*}
\MoveEqLeft[1] \Pr(Y_{j} = y, Y_{k} = y' \mid A_{xu}; \mathcal O) \\
&= \Pr\big(f(1,X_j,U_{j}) = y, f(0,X_k,U_{k}) = y' \mid A_{xu}; \mathcal O\big) \nonumber \\
&= \Pr\big(f(0,X_j,U_{j}) = y', f(1,X_k,U_{k}) = y \mid A_{xu}; \mathcal O\big) \nonumber \\
&= \left\{
\begin{array}{l}
1, \textrm{ when } (y, y') = \big(f(1,x,u), f(0,x,u)\big) \\
0, \textrm{ when } (y, y') \ne \big(f(1,x,u), f(0,x,u)\big). \nonumber
\end{array}\right.
\end{align*}
Thus, with this conditioning, the outcome is a deterministic function of the treatment assignment, and exchangeability applies trivially. This level of conditioning would be
required for identifying individual level causal effects, which is impossible in practice (the `fundamental problem of causal inference' as discussed by
\citealp{holland:1986}). The probabilistic condition is sufficient for identifying population-level effects. In the following section, we connect our concept of conditional exchangeability to Bayesian causal inference.

\section{Definition and estimation of causal contrasts}\label{section:causalcontrasts}

\subsection{Causal Contrasts Defined in Terms of Posterior Predictive Expectations}\label{section:definition}

As noted by \citet{greenland:2012}, causal inference can alternatively be formulated as a prediction problem or a missing data problem; the potential outcomes notation corresponds to the latter formulation. In the Bayesian framework, a causal contrast of interest may be naturally defined in terms of posterior predictive expectations for further exchangeable individuals under the hypothetical experimental setting already introduced above. We define the causal contrast of interest under the randomized setting in terms of the limits
\begin{align}\label{equation:causalcontrast}
\MoveEqLeft[1] \lim_{n \longrightarrow \infty} E[Y_{j} \mid Z_{j} = z, (w_i)_{i=1}^n; \mathcal E] \\
&- \lim_{n \longrightarrow \infty} E[Y_{k} \mid Z_{k} = z', (w_i)_{i=1}^n; \mathcal E], \nonumber
\end{align}
where $j \ne k > n$ and $(w_i)_{i=1}^n$ is a hypothetical exchangeable sequence under $\mathcal E$. We may consider such a contrast for arbitrary settings of the treatment indicators $z$ and $z'$, thus mimicking the classical `intervention' formulation of the causal contrast.  Note, however, that no special mathematical definitions or tools, other than those associated with fundamental exchangeability concepts, are required in this definition.

By de Finetti's representation theorem, the joint distribution of the data may be written
\settowidth{\templen}{$= \int_{\phi, \psi}$}
\begin{align}\label{formula:definetti}
\MoveEqLeft[0.5] \Pr\big((W_i)_{i=1}^n ; \mathcal E\big) \\
&= \int_\theta \prod_{i =1}^n P(W_{i} \mid \theta; \mathcal E) \,\textrm d Q(\theta ; \mathcal E) \nonumber \\
&= \int_{\phi, \psi} \prod_{i = 1}^n \left[P(Y_{i} \mid z_{i}, x_{i}, \phi; \mathcal E) P(X_{i} \mid \psi; \mathcal E) \right] \textrm d Q(\phi, \psi ; \mathcal E) \nonumber \\
&\hspace*{\templen}\times \int_{\gamma} \prod_{i = 1}^n P(Z_{i} \mid \gamma; \mathcal E) \,\textrm d Q(\gamma ; \mathcal E), \nonumber 
\end{align}
where $\theta = (\phi, \gamma, \psi)$ represents a partition of the joint parameter vector corresponding to the above factorization of the joint parameter-conditional distribution of $W_i = (Y_i, Z_i, X_i)$ in the second line, provided parameter $\Gamma$ is \emph{a priori} independent of the parameters $(\Phi, \Psi)$ \citep[cf.][p. 354--355]{gelman:2004}. Because all of these parameters are defined under $\mathcal E$, this independence follows from the factorization \eqref{eq: Expfactorize}, understanding the parameters as long-run summaries of the observable sequences. Now for any $j > n$ the expectations in (\ref{equation:causalcontrast}) may be written as
\begin{align}\label{formula:ppexpectation}
\MoveEqLeft[1] E[Y_{j} \mid z_{j}, (w_i)_{i=1}^n; \mathcal E] \\
&= \sum_{y_{j}, x_{j}} y_{j} \Pr(y_{j}, x_{j} \mid z_{j}, (w_i)_{i=1}^n; \mathcal E)
\nonumber \\
&= \frac{\sum_{y_{j}, x_{j}} y_{j} \int_{\phi, \psi} \prod_{i \in \{1, \ldots, n, j\}} L_i(\phi, \psi) \,\textrm d Q(\phi, \psi ; \mathcal E)}
{\sum_{y_{j}, x_{j}} \int_{\phi, \psi} \prod_{i \in \{1, \ldots, n, j\}} L_i(\phi, \psi) \,\textrm d Q(\phi, \psi ; \mathcal E)} \nonumber \\
&= \sum_{y_{j}, x_{j}} y_{j} \int_{\phi, \psi} L_j(\phi, \psi) \,\textrm d Q(\phi, \psi \mid (w_i)_{i=1}^n; \mathcal E), \nonumber
\end{align}
where $L_i(\phi, \psi) \equiv P(y_{i} \mid z_{i}, x_{i}, \phi; \mathcal E) P(x_{i} \mid \psi; \mathcal E)$. Here the terms involving parameters $\Gamma$ cancel out because $Z_i \independent X_i$ under $\mathcal E$ (and $\Gamma \independent (\Phi, \Psi)$); note that this would not hold under the observational setting $\mathcal O$.

If we further assume regularity conditions that allow interchanging the order of limit and integration, the limit of the above expectation becomes
\settowidth{\templen}{$= \sum_{x_{j}} \int_{\phi, \psi}$}
\begin{align}\label{formula:ppexpectationlimit}
\MoveEqLeft[1] \lim_{n \longrightarrow \infty} E[Y_{j} \mid z_{j}, (w_i)_{i=1}^n; \mathcal E] \\
&= \sum_{x_{j}} \int_{\phi, \psi} E[Y_{j} \mid z_{j}, x_{j}, \phi; \mathcal E] P(x_{j} \mid \psi; \mathcal E) \nonumber \\
&\hspace*{\templen}\times \delta_{\phi_0}(\phi) \delta_{\psi_0}(\psi) \, \textrm d\phi \, \textrm d\psi \nonumber \\
&= \sum_{x_{j}} E[Y_{j} \mid z_{j}, x_{j}, \phi_0; \mathcal E] P(x_{j} \mid \psi_0; \mathcal E), \nonumber
\end{align}
assuming that the posterior distribution converges to a degenerate distribution at the true parameter values $(\phi_0, \psi_0)$ \citep[cf.][p. 139]{vandervaart:1998}. The right-hand side here corresponds to the direct standardization/back-door formula, which was previously obtained informally as Equation \eqref{equation:ppdirectstandardization}. Because we interpret parameters as (unknown) functions of infinite sequences of observables (following \citealp[][p. 173]{bernardo:1994} and as per the definitions in Section \ref{section:introduction}), form \eqref{formula:ppexpectationlimit} motivates definition (\ref{equation:causalcontrast}) as the causal parameter of interest, as \eqref{formula:ppexpectationlimit}
does not depend on the prior $Q(\phi, \psi ; \mathcal E)$.

\subsection{Estimation Under the Observational Setting}\label{section:estimation}

\noindent \textbf{Identification:}
To estimate the causal contrast \eqref{equation:causalcontrast} defined under the experimental setting $\mathcal E$ based on data collected
under the observational setting $\mathcal O$, in \eqref{formula:ppexpectation} we have to make the substitutions $P(Y_{i} \mid z_{i}, x_{i}, \phi;
\mathcal E) = P(Y_{i} \mid z_{i}, x_{i}, \phi; \mathcal O)$ and $P(X_{i} \mid \psi; \mathcal E) = P(X_{i} \mid \psi; \mathcal O)$; the former corresponds
to the stability assumption, which in turn is implied by the infinite extension of the conditional exchangeability property \eqref{equation:spope}.
The latter can be taken to be true by definition, i.e., the standard population is chosen according to the observed covariate distribution. Under these
assumptions, parameters $\phi$ and $\psi$ have the \textbf{same} interpretation under both settings $\mathcal E$ and $\mathcal O$. With a given
observed realization $(w_i)_{i=1}^n$, this gives an \emph{estimator} for \eqref{formula:ppexpectationlimit} as
\begin{align}\label{equation:estimator}
\sum_{x_{j}} \int_{\phi, \psi} &E[Y_{j} \mid z_{j}, x_{j},  \phi; \mathcal O] P(x_{j} \mid \psi; \mathcal O) \\
&\times \,\textrm d Q(\phi, \psi \mid (w_i)_{i=1}^n; \mathcal O). \nonumber
\end{align}

We may also wish to state an identifiability condition in frequency-based terms. Because \eqref{equation:estimator} is taken to be the estimator of 
parameter \eqref{formula:ppexpectationlimit}, it is natural to require consistency, which we have if $\lim_{n \longrightarrow \infty} \textrm d Q(\phi,
\psi \mid (w_i)_{i=1}^n; \mathcal O) = \delta_{\phi_0}(\phi)\delta_{\psi_0}(\psi)$. In other words, the inferences will be \emph{unconfounded} if
\begin{align*}
\MoveEqLeft[1] \sum_{x_{j}} E[Y_{j} \mid z_{j}, x_{j}, \phi_0; \mathcal O] P(x_{j} \mid \psi_0; \mathcal O) \\
&= \lim_{n \longrightarrow \infty} E[Y_{j} \mid z_{j}, (w_i)_{i=1}^n; \mathcal E].
\end{align*}
A causal contrast could be defined alternatively in terms of potential outcome variables as $E[Y_{i}(1)] - E[Y_{i}(0)]$ . For unconfounded inferences, we could then require that
\begin{align}\label{equation:identifiability}
\sum_{x_i} E[Y_i \mid Z_i = z, x_i; \mathcal O] P(x_i ; \mathcal O) = E[Y_{i}(z)],
\end{align}
which follows from \eqref{equation:conditionalexchangeability2} \citep[e.g.,][p. 579]{hernan:2006}, and makes no explicit reference to the parametrization
of the problem.

\medskip

\noindent \textbf{Positivity:} To ensure that the conditional distributions above are well defined, we need an additional assumption known as \textit{positivity}, that is, absolute continuity of the two measures under $\mathcal E$ and $\mathcal O$ \citep[cf.][p. 196]{dawid:2010}, stated as $P(Z_{i} \mid x_{i}, \gamma; \mathcal E) \;\ll\; P(Z_{i} \mid x_{i}, \gamma; \mathcal O)$, which is equivalent to $P(Z_{i} \mid x_{i}, \gamma; \mathcal O) = 0 \;\Rightarrow\; P(Z_{i} \mid x_{i}, \gamma; \mathcal E) = 0$ or $P(Z_{i} \mid x_{i}, \gamma; \mathcal E) \neq 0 \;\Rightarrow\; P(Z_{i} \mid x_{i}, \gamma; \mathcal O) \neq 0$. In particular, if the treatment $Z_i$ depends deterministically on the covariates $X_i$, inference across the observational and experimental settings would not be possible.

Estimation of expectations \eqref{formula:ppexpectation} may be carried out using Monte Carlo integration by sampling from the posterior distribution of $(\Phi, \Psi)$. Because the distributions $P(Y_{i} \mid z_{i}, x_{i}, \phi; \mathcal O)$ and $P(X_{i} \mid \psi; \mathcal O)$ implied by the representation theorem are unknown, these have to be replaced with statistical models in practice. These models do not necessarily have to be parametric \citep[that is, having finite dimensional $\Phi$ and $\Psi$, cf.][p. 228]{bernardo:1994}, for instance, we would usually model $P(X_{i} \mid \psi; \mathcal O)$ with the empirical distribution of $X_i$; however, in practice, the curse of dimensionality limits the use of nonparametric specifications for the outcome model, and dimension-reducing modeling assumptions will become a necessity. When finite dimensional parametrizations are used, model misspecification becomes a potential issue. In particular, one may lose the important property of valid inferences under the null hypothesis of no treatment effect, which will be elaborated on in the following section.

\section{Longitudinal setting: Exchangeability and sequential randomization}\label{section:nullparadox}

Having established the Bayesian formulation of causal inference in point treatment settings, we now seek to extend this reasoning to the longitudinal
case, where confounding structures may be more complex.  For simplicity, we consider the two time-point case and contend that the extension to multiple
time-points follows straightforwardly. Consider now the slightly more complicated setting in the DAG in the right panel of Figure \ref{figure:dag3}, labeled by $\mathcal O$, adapted from \citet{robins:1997}, where the design variables $Z_{1i}$ and $Z_{2i}$ represent the treatment assignment to initiate or receive a particular dose of antiretroviral medication starting at baseline and at a subsequent re-examination, respectively, for individual $i$. Further, let $X_i$ represent observed anemia status at the re-examination, and $Y_i$ an HIV viral load
outcome, measured at the end of follow-up after sufficient time has passed from the re-examination. Latent variable $U_i$ again represents the underlying
immune function of individual $i$, which is expected to be a determinant of both $X_i$ and $Y_i$. Here $X_i$ being influenced by earlier treatment introduces treatment-confounder feedback \citep[][p. 550]{robins:2000}, which makes the judgment of exchangeability somewhat more involved. For the causal exchangeability considerations, we take the intermediate variable and outcome to be determined by structural models $X_i = g(Z_{1i},U_i)$ and $Y_i = f(Z_{1i}, Z_{2i}, X_i, U_i)$.

\tikzset{
    -Latex,auto,node distance =1 cm and 1 cm,semithick,
    state/.style ={circle, draw, minimum width = 0.7 cm},
    box/.style ={rectangle, draw, minimum width = 0.7 cm, fill=lightgray},
    point/.style = {circle, draw, inner sep=0.08cm,fill,node contents={}},
    bidirected/.style={Latex-Latex,dashed},
    el/.style = {inner sep=3pt, align=left, sloped}
}
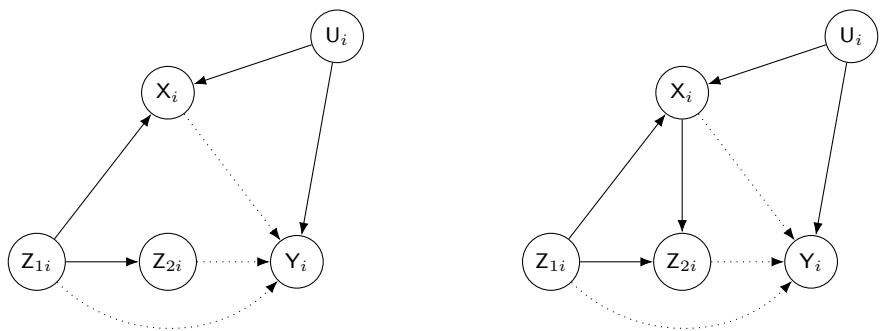
\begin{figure}[!hb]
\centering
\begin{tikzpicture}[scale=1.5]
    \node[state] (z2) at (0,0) {$\textsf{Z}_{2i}$};
    \node[state] (x) at (0,1.5) {$\textsf{X}_i$};
    \node[state] (u) at (1.5,2) {$\textsf{U}_i$};

    \node (e) at (-1.5,2) {$\mathcal E$};

    \node[state] (y) [right =of z2] {$\textsf{Y}_i$};
    \node[state] (z1) [left =of z2] {$\textsf{Z}_{1i}$};

    \path (z2) edge[dotted] (y);
    \path (z1) edge (x);
    \path (z1) edge (z2);
    \path (x) edge[dotted] (y);
    \path (u) edge (y);

    \path (z1) edge[dotted, bend right=45] (y);

    \path (u) edge[solid] (x);

\end{tikzpicture}
\hspace{0.5in}
\begin{tikzpicture}[scale=1.5]
    \node[state] (z2) at (0,0) {$\textsf{Z}_{2i}$};
    \node[state] (x) at (0,1.5) {$\textsf{X}_i$};
    \node[state] (u) at (1.5,2) {$\textsf{U}_i$};

    \node (o) at (-1.5,2) {$\mathcal O$};

    \node[state] (y) [right =of z2] {$\textsf{Y}_i$};
    \node[state] (z1) [left =of z2] {$\textsf{Z}_{1i}$};

    \path (z2) edge[dotted] (y);
    \path (z1) edge (x);
    \path (z1) edge (z2);
    \path (x) edge[dotted] (y);
    \path (x) edge (z2);
    \path (u) edge (y);

    \path (z1) edge[dotted, bend right=45] (y);

    \path (u) edge[solid] (x);

\end{tikzpicture}
\caption{Left panel: DAG depicting the randomized longitudinal setting labeled by $\mathcal E$. The dashed arrows are absent under the null hypothesis of no treatment effect in the presence of treatment-confounder feedback. Note that the null hypothesis also holds under an alternative DAG, where the arrow $Z_{1i} \longrightarrow X_i$ is omitted, and the dotted arrow $X_i \longrightarrow Y_i$ may be present. Right panel: DAG depicting the observational longitudinal setting labeled by $\mathcal O$. This DAG differs from that on the left by the arrow $X_i \longrightarrow Z_{2i}$}\label{figure:dag3}
\end{figure}

The principal source of difficulty is represented by the latent variable $U_i$. To consider its implications for inference, we first define the causal contrast of interest in terms of a randomized setting labeled by $\mathcal E$ depicted in the left panel of Figure \ref{figure:dag3}. We may now define the causal contrast of interest as
\begin{align*}
\MoveEqLeft[1] \lim_{n \longrightarrow \infty} E[Y_{j} \mid Z_{1j} = z_{1j}, Z_{2j} = z_{2j}, (w_i)_{i=1}^n; \mathcal E] \\
&- \lim_{n \longrightarrow \infty} E[Y_{k} \mid Z_{1k} = z_{1k}, Z_{2k} = z_{2k}, (w_i)_{i=1}^n; \mathcal E],
\end{align*}
where $j \ne k > n$. The expectations here can be represented alternatively as
\begin{align}
\label{formula:ppexpectationlimit2}
\MoveEqLeft[1] \lim_{n \longrightarrow \infty} E[Y_{i} \mid z_{1i}, z_{2i}, (w_i)_{i=1}^n; \mathcal E] \\
&= E[Y_{i} \mid z_{1i}, z_{2i}, \varphi_0; \mathcal E] \nonumber
\end{align}
or
\begin{align}
\label{formula:ppexpectationlimit3}
\MoveEqLeft[1] \lim_{n \longrightarrow \infty} E[Y_{i} \mid z_{1i}, z_{2i}, (w_i)_{i=1}^n; \mathcal E] \\
&= \sum_{x_{i}} E[Y_{i} \mid z_{1i}, z_{2i}, x_{i}, \phi_0^*; \mathcal E] P(x_{i} \mid z_{1i}, \psi_0^*; \mathcal E) \nonumber
\end{align}
or finally
\settowidth{\templen}{$= \sum_{x_{i}} \int_{u_{i}}$}
\begin{align}\label{formula:ppexpectationlimit4}
\MoveEqLeft[1] \lim_{n \longrightarrow \infty} E[Y_{i} \mid z_{1i}, z_{2i}, (w_i)_{i=1}^n; \mathcal E] \\
&= \sum_{x_{i}} \int_{u_{i}} E[Y_{i} \mid z_{1i}, z_{2i}, x_{i},  u_{i}, \phi_0^\dagger; \mathcal E] \nonumber \\
& \hspace*{\templen}\times P(x_{i} \mid z_{1i}, u_{i}, \psi_0^\dagger; \mathcal E) P(\textrm d u_{i} \mid \eta_0^\dagger; \mathcal E). \nonumber 
\end{align}
Note the different parameters $\varphi, (\phi^*,\psi^*),$ and $(\phi^\dagger,\psi^\dagger,\eta^\dagger)$ in the three representations. The parametrization in \eqref{formula:ppexpectationlimit4} corresponds to the data generating mechanism, the parameters of which are determined by the representation
for infinitely exchangeable random vectors $( Y_{i}, Z_{1i}, Z_{2i}, X_{i}, U_i)$, whereas the parameters that appear in
\eqref{formula:ppexpectationlimit2} and \eqref{formula:ppexpectationlimit3} are consequences of the joint model obtained by marginalization.

As was done in Section \ref{section:individuallevel}, we consider for simplicity binary or dichotomized treatments and consider the comparability
of groups selected to have a given treatment assignment configuration. The exchangeability with respect to the intermediate variable $X_i$ can be
established as before. For the outcome $Y_i$, we consider exchangeability separately at the time of each treatment. The groups being compared have
the treatment assignments
$(Z_{1i} = 1, Z_{2i} = 1)$,
$(Z_{1i} = 1, Z_{2i} = 0)$,
$(Z_{1i} = 0, Z_{2i} = 1)$, and
$(Z_{1i} = 0, Z_{2i} = 0)$. We note that the parameters $\varphi$ in the outcome model $P(Y_i \mid z_{1i}, z_{2i}, \varphi; \mathcal E)$ corresponding
to parametrization \eqref{formula:ppexpectationlimit2} would not be estimatable under the observational setting $\mathcal O$. For
instance, at the second time point, the outcomes of individuals $j$ and $k$ with opposite treatment assignments would not be exchangeable (those assigned
to treatment at the second interval are likely to have better underlying immune function status than those not assigned to treatment, with the second
assignment depending on $X_i$), that is, we do not have that
\settowidth{\templen}{$= \Pr\big($}
\begin{align*}
\MoveEqLeft[1] \Pr(Y_{j} = y, Y_{k} = y' \mid A; \mathcal O) \\
&= \Pr\big(f(1,1,g(1,U_j),U_j) = y, \\
&\hspace*{\templen} f(1,0,g(1,U_k),U_k) = y' \mid A; \mathcal O\big) \\ \nonumber
&= \Pr\big(f(1,1,X_j,U_j) = y, f(1,0,X_k,U_k) = y' \mid A; \mathcal O\big) \\ \nonumber
&= \Pr\big(f(1,0,X_j,U_j) = y', f(1,1,X_k,U_k) = y \mid A; \mathcal O\big), \nonumber
\end{align*}
where $A \equiv \{Z_{1j} = Z_{1k} = 1, Z_{2j} = 1, Z_{2k} = 0\}$.

Instead, we can adopt parametrization \eqref{formula:ppexpectationlimit3} and model the conditional distributions $P(Y_{i} \mid z_{1i}, z_{2i},x_{i}, 
\phi^*; \mathcal O)$ and $P(X_i \mid z_{1i}, \psi^*; \mathcal E)$. Now based on Figure \ref{figure:dag3} we have that $Z_{1i} \independent U_i$ and $Z_{2i} \independent U_i \mid (Z_{1i}, X_i)$ under $\mathcal O$, which together imply the sequential randomization condition discussed by e.g., \citet[][p. 200]{dawid:2010}, or stability $P(Y_i \mid Z_{1i}, Z_{2i}, X_i; \mathcal E) = P(Y_i \mid Z_{1i}, Z_{2i}, X_i; \mathcal O)$ and $P(X_i \mid Z_{1i}; \mathcal E) = P(X_i \mid Z_{1i}; \mathcal O)$. Stability would be sufficient to ensure non-parametric identification of the marginal causal contrast because
\settowidth{\templen}{$= \sum_{x_i} \int_{u_i}$}
\begin{align*}
\MoveEqLeft[1] P(Y_i \mid z_{1i}, z_{2i}; \mathcal E) \\
&= \sum_{x_i} \int_{u_i} \frac{P(Y_i, Z_{1i} = z_{1i}, Z_{2i} = z_{2i}, x_i, \textrm du_i ; \mathcal E)}{P(Z_{1i} = z_{1i}, Z_{2i} = z_{2i} ; \mathcal E)} \\
&= \sum_{x_i} \int_{u_i} P(Y_i \mid z_{1i}, z_{2i}, x_i, u_i; \mathcal E)P(x_i \mid z_{1i}, u_i; \mathcal E) \\
& \hspace*{\templen}\times P(\textrm du_i \mid z_{1i}; \mathcal E) \\
&= \sum_{x_i} \int_{u_i} P(Y_i \mid z_{1i}, z_{2i}, x_i, u_i; \mathcal E)P(x_i \mid z_{1i}; \mathcal E) \\
& \hspace*{\templen}\times P(\textrm du_i \mid z_{1i}, z_{2i}, x_i; \mathcal E) \\
&= \sum_{x_i} \int_{u_i} P(Y_i, \textrm du_i \mid z_{1i}, z_{2i}, x_i; \mathcal E) P(x_i \mid z_{1i}; \mathcal E) \\
&= \sum_{x_i} P(Y_i \mid z_{1i}, z_{2i}, x_i; \mathcal E) P(x_i \mid z_{1i}; \mathcal E) \\
&= \sum_{x_i} P(Y_i \mid z_{1i}, z_{2i}, x_i; \mathcal O) P(x_i \mid z_{1i}; \mathcal O).
\end{align*}

However, under the longitudinal setting introducing stratification by $X_i$ does not restore the conditional exchangeability of all the groups being
compared.  We now do have exchangeability between individuals $j$ and $k$ with opposing treatment assignments at the second time point, that is,
\begin{align}\label{equation:conditionalexchangeability5}
\MoveEqLeft[1] \Pr(Y_{j} = y, Y_{k} = y' \mid A_x ; \mathcal O) \\
&= \Pr\big(f(1,1,x,U_j) = y, f(1,0,x,U_k) = y' \mid A_x ; \mathcal O\big) \nonumber \\
&= \Pr\big(f(1,0,x,U_j) = y', f(1,1,x,U_k) = y \mid A_x ; \mathcal O\big), \nonumber
\end{align}
where $A_x \equiv \{Z_{1j} = Z_{1k} = 1, Z_{2j} = 1, Z_{2k} = 0, X_{j} = X_{k} = x\}$. However, when comparing individuals with opposite treatment
assignments at the first time point, the conditional exchangeability condition
\begin{align}\label{equation:conditionalexchangeability4}
\MoveEqLeft[1] \Pr(Y_{j} = y, Y_{k} = y' \mid A_x ; \mathcal O) \\
&= \Pr\big(f(1,0,x,U_j) = y, f(0,0,x,U_k) = y' \mid A_x ; \mathcal O\big)  \nonumber \\
&= \Pr\big(f(0,0,x,U_j) = y', f(1,0,x,U_k) = y \mid A_x ; \mathcal O\big), \nonumber
\end{align}
where $A_x \equiv \{Z_{1j} = 1, Z_{1k} = 0, Z_{2j} =Z_{2k} = 0, X_{j} = X_{k} = x\}$, does \textbf{not} hold because the prior information we have on the relationships between the variables indicates, for example, that those without anemia and assigned to treatment at the first interval are likely to have better immune function status than those without anemia and no treatment at the first interval because initiation of the treatment is in itself a cause of anemia. This would be the case also if the groups being compared had been formed under the completely randomized setting $\mathcal E$, even though the groups would be exchangeable without the stratification. In the causal inference literature this phenomenon has been called \textit{collider stratification bias} \citep[e.g.,][]{greenland:2003}, \textit{Berkson's bias} or merely \textit{selection bias}; as demonstrated, it can equally well be understood as lack of conditional exchangeability of the groups being compared in terms of their pre-treatment characteristics. Exchangeability does hold matching on the initial treatment assignment $Z_{i1}$, but this would not allow estimation of the effect of $Z_{i1}$. The non-exchangeability of the groups not matched with respect to the initial treatment assignment is illustrated in the numerical example presented in Supplementary Appendix B.

The lack of conditional exchangeability corresponding to \eqref{equation:conditionalexchangeability4} implies that the parameters $\phi^*$ in the
conditional probability model $P(Y_{i} \mid  z_{1i}, z_{2i}, x_{i}, \phi^*; \mathcal O)$ characterizing the association between $Y_i$ and $Z_{2i}$
would not have a causal interpretation, and thus a modeling strategy based on finite-dimensional parametrization $(\phi^*, \psi^*)$ might not be
successful; without an appropriate parametrization of the problem, we may lose the important property of valid inferences under the null
hypothesis of no treatment effect, which gives rise to the so-called \textit{null paradox} (e.g., \citealp[][p. 411-412]{robins:1997},
\citealp{vansteelandt:2012}, p. 11; \citealp{dawid:2010}, p. 224).

The conditional exchangeability condition (\ref{equation:conditionalexchangeability4}) relates to the stronger conditional independence condition $(Z_{1i}, Z_{2i}) \independent U_i \mid X_i$ required for identification of controlled direct effects \citep[e.g.,][]{greenland:1992,vanderweele:2009b}. This does not hold under the setting of Figure \ref{figure:dag3}, but exchangeability could be restored by introducing further conditioning on $U_i$, which implies that \eqref{formula:ppexpectationlimit4} would be the correct causal parametrization. However, because $U_i$ is unobserved, the use of such parametrization in practice would introduce new identifiability problems. The null-robust reparametrization of the problem, as suggested by \citet[][p. 415-416]{robins:1997} might be one way to proceed.

Regardless of the issues related to finite-dimensional parametrizations, we note that a connection between conditional exchangeability statements and the stability property is still preserved in the longitudinal setting. As we have noted above, sequential randomization is sufficient for stability, and assuming conditional exchangeability under permutations of both treatment assignments $Z_{1i}$ and $Z_{2i}$ is unnecessarily strong for non-parametric identifiability of the problem. If we assume the infinite extension of exchangeability property \eqref{equation:conditionalexchangeability5} with respect to permutations $Z_{2i}$ at fixed levels of $Z_{1i}$, we note that at the second timepoint $Z_{1i}$ has the same role as the observed confounders $X_i$. We can then use the same arguments as in Sections \ref{sec:Ex_ind} and \ref{section:individuallevel} to find that $Z_{2i} \independent f(z_1, z_2, x, U_i) \mid (Z_{1i} = z_1, X_i = x)$ under both $\mathcal E$ and $\mathcal O$. This corresponds to the second condition of sequential randomization and can be used to further obtain $\Pr(Y_i = y \mid Z_{i1} = z_1, Z_{i2} = z_2, X_i = x; \mathcal E) = \Pr(f(z_1,z_2,x,U_i) = y \mid Z_{i1} = z_1, Z_{i2} = z_2, X_i = x; \mathcal E) = \Pr(f(z_1,z_2,x,U_i) = y \mid Z_{i1} = z_1, X_i = x; \mathcal E)$. Here
\begin{align*}
\MoveEqLeft[1] \Pr(f(z_1,z_2,x,U_i) = y \mid Z_{i1} = z_1, X_i = x; \mathcal E)  \\
&= \frac{\Pr(f(z_1,z_2,x,U_i) = y, g(z_1,U_i) = x \mid Z_{i1} = z_1; \mathcal E)}{\Pr(g(z_1,U_i) = x \mid Z_{i1} = z_1; \mathcal E)}.
\end{align*}
Thus, if we have $(f(z_1,z_2,g(z_1,U_i),U_i), g(z_1,U_i)) \independent Z_{1i}$ (which in turn implies that $g(z_1,U_i) \independent Z_{1i}$), under the usual assumption that the distribution of the baseline characteristics is the same under $\mathcal E$ and $\mathcal O$, we can get that 
\begin{align*}
\MoveEqLeft[1] \Pr(f(z_1,z_2,x,U_i) = y \mid Z_{i1} = z_1, X_i = x; \mathcal E) \\
&= \Pr(f(z_1,z_2,x,U_i) \mid Z_{i1} = z_1, X_i = x; \mathcal O) \\
&= \Pr(f(z_1,z_2,x,U_i) \mid Z_{i1} = z_1, Z_{i2} = z_2, X_i = x; \mathcal O) \\
&= \Pr(Y_i \mid Z_{i1} = z_1, Z_{i2} = z_2, X_i = x; \mathcal O). 
\end{align*}
Using similar arguments as before, these properties also apply in the i.i.d. distribution, implying the stability property for the outcome distribution. The first sequential randomization condition $Z_{1i} \independent U_i$ would be sufficient for the required independence, but it can also be obtained from the infinite joint exchangeability property for sequences of $f(z_1,z_2,g(z_1,U_i),U_i)$ and $g(z_1,U_i)$ conditional on $Z_{1i}$. Thus, we contend that while obtaining identifying conditions for causal effects based on conditional exchangeability statements is more cumbersome in the presence of treatment-confounder feedback, it appears to be possible. We also note that the required identifying conditions correspond to $(Y_{i}(z_1,z_2), X_{i}(z_1)) \independent Z_{1i}$ and $Y_{i}(z_1, z_2) \independent Z_{2i} \mid (Z_{1i}, X_i)$ expressed in terms of potential outcome variables if we take $Y_{i}(z_1, z_2) = f(z_1,z_2,g(z_1,U_i),U_i)$ and $X_{i}(z_1) = g(z_1,U_i)$, i.e., the treatment assignments are independent of future potential outcomes and intermediate variables conditional on observed past \citep[e.g.,][]{chakraborty2014dynamic}.

\section{Discussion}\label{section:discussion}

We have demonstrated that the notion of exchangeability as a probabilistic symmetry property can indeed serve as as a basis of a causal model, as was originally suggested by \citet{lindley:1981}. That exchangeability can be formulated as an ignorability assumption, and that marginal causal contrasts can be naturally defined in terms of limits of posterior predictive expectations for further, yet unobserved, exchangeable individuals, has not been appreciated in the causal inference literature. We do not claim that the interpretation of exchangeability as a causal model would have important practical advantages over alternative causal models; the preference for a particular causal model as the notational system is largely a matter of taste and convention. In particular, the identifying conditions required for inferences were equivalent to corresponding conditions stated in terms of potential outcomes. However, the proposed framework links causality more closely to model parameters and does enable a more natural incorporation of causal reasoning into the fully probabilistic Bayesian framework, in the sense that no concepts external to de Finetti's system are necessary.

We demonstrated a connection between conditional exchangeability statements and causal interpretation of parameters in statistical models. However, in the longitudinal setting of Section \ref{section:nullparadox}, the connection between the conditional exchangeability properties corresponding to the model components and identifying conditions for marginal causal contrasts defined without reference to statistical models becomes less direct. In particular, in situations where conditioning on intermediate variables opens backdoor paths between treatments and the outcome, component models may not be interpretable, while the marginal causal effects may still be identifiable. Alternative inference methods exist that can identify the causal contrast under the sequential randomization condition and with fewer parametric modeling assumptions; consider, for example, marginal structural models estimated using inverse probability of treatment weighting \citep{robins:2000,hernan:2001}. Nonetheless, exchangeability judgments may warn us of a situation where null paradox type model misspecification issues are likely to arise.  Proper understanding of the problem and the possible solutions are especially important given the recent renewed interest in the parametric $g$-computation formula \citep[e.g.,][]{taubman:2009,westreich:2012,keil2014parametric,jain2016smoking,bijlsma2017unemployment,neophytou2019diesel,Shahn:2019}. The issues related to finite dimensional parametrizations also motivate further research into semi-parametric Bayesian inference procedures, which would allow direct parametrization of marginal causal effects while avoiding specifications of some of the likelihood components \citep[cf.][]{saarela:2015,
saarela2016bayesian}.

Throughout, we assumed a functional relationship between the outcome and its determinants, with the function $f(z, X_i, U_i)$ understood as the equivalent of the potential outcome $Y(z)$. This notation allows us to decouple the observed, potentially informative, treatment assignment from the intervention in the exchangeability judgments when considering switching the treatment of the units. The assumed deterministic relationship may not be a serious limitation, as $U_i$ could always be thought to include the remaining (unobserved) determinants of the outcome. However, a reviewer points out that the present framework could potentially be modified to allow for stochastic dependency of $Y$ on $(Z, X, U)$ by introducing separate notation for the intended/assigned treatment $Z$ and intervention to administer treatment $\widehat Z$. One could then consider exchangeability statements of the type 
\begin{align*}
\MoveEqLeft[1] \Pr(Y_j = y; Y_k = y' \mid Z_j = z, Z_k = z'; \widehat Z_j = z,  Z_k = z')\\
&= \Pr(Y_j = y'; Y_k = y \mid Z_j = z, Z_k = z'; \widehat Z_j = z',  Z_k = z),
\end{align*}
conditional on both the assignment and the intervention (which may be different). This has a similar interpretation as \eqref{equation:populationexchangeabilitye} but does not require introducing the functional relationship for the outcome. We leave it as further work to study whether the presented framework can be adapted accordingly to obtain the same results.

\begin{supplement}
\stitle{Supplementary Appendix A}
\sdescription{Includes Bayesian DAGs.}
\end{supplement}

\begin{supplement}
\stitle{Supplementary Appendix B}
\sdescription{Includes numerical examples.}
\end{supplement}

\bibliographystyle{imsart-nameyear}
\bibliography{refs}

\section*{Appendix: Likelihood construction under dichotomous outcomes}

Suppose that $Y_i \in \{0,1\}$ is an outcome event indicator or dichotomized continuous or count outcome and the subsequences of these indicator variables for treated units $Z_i = 1$ and untreated units $Z_i = 0$ are separately infinitely exchangeable, that is, we have partial exchangeability
\settowidth{\templen}{$= \Pr \Bigg($}
\begin{align*}
\MoveEqLeft[1] \Pr \left(\bigcap_{i: z_i = 1} (Y_{i} = y_{i}), \bigcap_{i: z_i = 0} (Y_i = y_{i}) \bigg | \bigcap_{i=1}^n (Z_i = z_i); \mathcal E  \right) \nonumber \\
&= \Pr \Bigg(\bigcap_{i: z_i = 1} (Y_i = y_{\rho_1(i)}), \\
&\hspace*{\templen} \bigcap_{i: z_i = 0} (Y_i = y_{\rho_0(i)}) \bigg | \bigcap_{i=1}^n (Z_i = z_i); \mathcal E \Bigg)
\end{align*}
for any permutations $\rho_1$ and $\rho_0$ of the subsequences. If in addition, we assume that the treated event count and untreated event count are
sufficient statistics, by Proposition 4.18 of \citet{bernardo:1994}, for each pair of treated and untreated units we have that
\settowidth{\templen}{$= \int_{[0,1]^2}$}
\begin{align}\label{lik}
\MoveEqLeft[0.5] \Pr(Y_j = y, Y_k = y' \mid Z_j = 1, Z_k = 0; \mathcal E) \\
&= \int_{[0,1]^2} \Pr(Y_j = y, Y_k = y' \mid Z_j = 1, Z_k = 0, \phi_1 , \phi_0; \mathcal E) \nonumber \\
&\hspace*{\templen}\times \,\textrm d Q(\phi ; \mathcal E) \nonumber \\
&= \int_{[0,1]^2} \phi_1^{y} (1-\phi_1)^{1-y} \phi_0^{y'} (1-\phi_0)^{1-y'} \,\textrm d Q(\phi ; \mathcal E), \nonumber 
\end{align}
where $\phi = (\phi_0, \phi_1)$,
\settowidth{\templen}{$= \displaystyle \lim_{n \longrightarrow \infty} \Pr\Bigg($}
\settowidth{\templennew}{$\Pr\Bigg($}
\begin{align*}
\MoveEqLeft[0.5] Q(\phi; \mathcal E) \\
&= \lim_{n \longrightarrow \infty} \Pr\Bigg( \frac{\sum_{i=1}^n z_i Y_i}{\sum_{i=1}^n z_i} \le \phi_1, \\
&\hspace*{\templen} \frac{\sum_{i=1}^n (1-z_i) Y_i}{\sum_{i=1}^n (1-z_i)} \le \phi_0 \bigg | \bigcap_{i=1}^n (Z_i = z_i) \Bigg) \\
&= \lim_{n \longrightarrow \infty} \\
&\quad\Pr\Bigg( \frac{\sum_{i=1}^n z_i f(1,X_i,U_i)}{\sum_{i=1}^n z_i} \le \phi_1, \\
&\quad\hspace*{\templennew}\frac{\sum_{i=1}^n (1-z_i) f(0,X_i,U_i)}{\sum_{i=1}^n (1-z_i)} \le \phi_0 \bigg |  \bigcap_{i=1}^n (Z_i = z_i) \Bigg),
\end{align*}
and
\begin{align*}
\phi_1 &= \lim_{n \longrightarrow \infty} \frac{\sum_{i=1}^n z_i Y_i}{\sum_{i=1}^n z_i} \\
&= \lim_{n \longrightarrow \infty} \frac{\sum_{i=1}^n z_i f(1,X_i,U_i)}{\sum_{i=1}^n z_i} 
\end{align*}
and
\begin{align*}
\phi_0 &= \lim_{n \longrightarrow \infty} \frac{\sum_{i=1}^n (1-z_i) Y_i}{\sum_{i=1}^n (1-z_i)} \\
&= \lim_{n \longrightarrow \infty} \frac{\sum_{i=1}^n (1-z_i) f(0,X_i,U_i)}{\sum_{i=1}^n (1-z_i)}.
\end{align*}

As noted before, for contrasts of the treated and untreated outcome event frequencies such as $\phi_1 - \phi_0$ or $\phi_1/\phi_0$ to have any causal interpretation, we need further assumptions in addition to the partial exchangeability. From the general exchangeability, it follows that
\begin{align*}
\MoveEqLeft[1] \Pr(Y_{j} = y, Y_{k} = y' \mid Z_{j} = 1, Z_{k} = 0; \mathcal E) \\
&= \frac{\Pr(Y_{j} = y, Y_{k} = y', Z_{j} = 1, Z_{k} = 0; \mathcal E)}{\Pr(Z_{j} = 1, Z_{k} = 0; \mathcal E)} \\
&= \frac{\Pr(Y_{j} = y', Y_{k} = y, Z_{j} = 0, Z_{k} = 1; \mathcal E)}{\Pr(Z_{j} = 0, Z_{k} = 1; \mathcal E)} \\
&= \Pr(Y_{j} = y', Y_{k} = y \mid Z_{j} = 0, Z_{k} = 1; \mathcal E).
\end{align*}
Further, by the structural model and \eqref{equation:populationexchangeabilitye}, we have that
\settowidth{\templen}{$= \Pr\big($}
\begin{align*}
\MoveEqLeft[1] \Pr(Y_{j} = y', Y_{k} = y \mid Z_{j} = 0, Z_{k} = 1; \mathcal E) \\
&= \Pr\big(f(1,X_k,U_k) = y, \\
& \hspace*{\templen} f(0,X_j,U_j) = y' \mid Z_{j} = 0, Z_{k} = 1; \mathcal E\big) \nonumber \\
&= \Pr\big(f(1,X_j,U_j) = y, \\
& \hspace*{\templen} f(0,X_k,U_k) = y' \mid Z_{j} = 0, Z_{k} = 1; \mathcal E\big),
\end{align*}
so that
\settowidth{\templen}{$\Pr\big($}
\begin{align*}
&\Pr\big(f(1,X_j,U_j) = y, \\
&\hspace*{\templen} f(0,X_k,U_k) = y' \mid Z_{j} = 1, Z_{k} = 0; \mathcal E\big) \\
&\quad= \Pr\big(f(1,X_j,U_j) = y, \\
&\quad\quad\hspace*{\templen} f(0,X_k,U_k) = y' \mid Z_{j} = 0, Z_{k} = 1; \mathcal E\big).
\end{align*}
This indicates that under the added conditional exchangeability assumption, the joint distribution of $f(1,X_j,U_j)$ and $f(0,X_k,U_k)$ does not depend
on which one of $j$ and $k$ was actually assigned the treatment. More generally, we have that
\begin{align}\label{independence}
\MoveEqLeft[1] \Pr \left(\bigcap_{i=1}^n (f(z_i,X_i,U_i) = y_{i}) \bigg | \bigcap_{i=1}^n (Z_{i} = z_i); \mathcal E  \right) \\
&= \Pr \left(\bigcap_{i=1}^n (f(z_i,X_{i},U_{i}) = y_{i}) \bigg | \bigcap_{i=1}^n (Z_{\rho(i)} = z_{i}); \mathcal E  \right).\nonumber 
\end{align}
This indicates that the limiting distribution $Q(\phi; \mathcal E)$ and the limits $\phi_1$ and $\phi_0$ do not depend on the actual treatment
assignment. We also note that from \eqref{exinf} it follows that
\settowidth{\templen}{$\Pr \Bigg($}
\begin{align*}
&\Pr \Bigg(\bigcap_{i: z_i = 1} (f(1,X_i,U_i) = y_{i}), \\
&\hspace*{\templen} \bigcap_{i: z_i = 0} (f(0,X_i,U_i) = y_{i}) \bigg | \bigcap_{i=1}^n (Z_{i} = z_i); \mathcal E  \Bigg) \nonumber \\
&\quad= \Pr \Bigg(\bigcap_{i: z_i = 1} (f(1,X_i,U_i) = y_{\rho_1(i)}), \\
&\quad\quad\hspace*{\templen} \bigcap_{i: z_i = 0} (f(0,X_i,U_i) = y_{\rho_0(i)}) \bigg | \bigcap_{i=1}^n (Z_{i} = z_i); \mathcal E  \Bigg) \nonumber
\end{align*}
for any permutations $\rho_1$ and $\rho_0$ of the treated and untreated subsequences. This means that partial exchangeability still applies to the sequences
of under treatment random variables $f(1,X_i,U_i)$ and without treatment random variables $f(0,X_i,U_i)$, enabling application of Proposition 4.18 of \citet{bernardo:1994}
directly to the joint distribution of these. Thus, we conclude that the likelihood is given by
\settowidth{\templen}{$=\Pr \big($}
\begin{align*}
\MoveEqLeft[1] \Pr(Y_j = y, Y_k = y' \mid Z_j = 1, Z_k = 0, \phi_1 , \phi_0; \mathcal E) \\
&= \Pr\big(f(1,X_j,U_j) = y, \\
& \hspace*{\templen} f(0,X_k,U_k) = y' \mid Z_j = 1, Z_k = 0, \phi_1 , \phi_0; \mathcal E\big) \\
&= \Pr\big(f(1,X_k,U_k) = y, \\
& \hspace*{\templen} f(0,X_j,U_j) = y' \mid Z_j = 1, Z_k = 0, \phi_1, \phi_0; \mathcal E\big),
\end{align*}
so symmetry property \eqref{equation:populationexchangeabilitye} is preserved conditional on the parameters.

\end{document}